\documentclass[letterpaper, 10 pt, conference]{ieeeconf}  % Comment this line out
\IEEEoverridecommandlockouts                              % This command is only
\usepackage{amsmath} % assumes amsmath package installed
\usepackage{amssymb}  % assumes amsmath package installed
\usepackage{mathtools}
\usepackage{algorithm}
\usepackage{algpseudocode}

\usepackage{graphicx}
\usepackage{subcaption}
\graphicspath{ {./graphics/} }
\usepackage{booktabs}
\usepackage{xcolor}

\title{\LARGE \bf
Contingency-constrained economic dispatch \\ with safe reinforcement learning
}

\author{Michael Eichelbeck, Hannah Markgraf, and Matthias Althoff%
\thanks{This work was partially supported by the German Research Foundation (AL 1185/9-1) and the Bavarian Research Foundation project STROM (Energy - Sector coupling and microgrids). The authors are with the Department of Informatics, Technical University
of Munich, Boltzmannstr. 3, 85748 Garching, Germany.
E-mail: {\tt\small \{eic, mhan, althoff\}@in.tum.de}}%
}

\begin{document}

\maketitle
\thispagestyle{empty}
\pagestyle{empty}

%%%%%%%%%%%%%%%%%%%%%%%%%%%%%%%%%%%%%%%%%%%%%%%%%%%%%%%%%%%%%%%%%%%%%%%%%%%%%%%%
\begin{abstract}

Future power systems will rely heavily on micro grids with a high share of decentralised renewable energy sources and energy storage systems. 
The high complexity and uncertainty in this context might make conventional power dispatch strategies infeasible.
Reinforcement-learning-based (RL) controllers can address this challenge, however, cannot themselves provide safety guarantees preventing their deployment in practice. To overcome this limitation, we propose a formally validated RL controller for economic dispatch.
We extend conventional constraints by a time-dependent constraint encoding the islanding contingency. The contingency constraint is computed using set-based backwards reachability analysis and actions of the RL agent are verified through a safety layer. Unsafe actions are projected into the safe action space while leveraging constrained zonotope set representations for computational efficiency. 
The developed approach is demonstrated on a residential use case using real-world measurements.

\end{abstract}

%%%%%%%%%%%%%%%%%%%%%%%%%%%%%%%%%%%%%%%%%%%%%%%%%%%%%%%%%%%%%%%%%%%%%%%%%%%%%%%%
\section{INTRODUCTION}

We investigate economic dispatch of micro grids with a high share of decentralised renewable sources. 
The central challenge of operating such grids is the unreliable generative capacity of renewable energy sources, which has to be mitigated by energy storage systems. The goal of economic dispatch is to determine the optimal power output of generators and energy storage systems such that power demand is satisfied and the overall operational cost is minimal. 

Existing works propose, for example, mixed-integer programming \cite{malysz_optimal_2014}, \cite{macedo_optimal_2015}, distributed primal-dual optimization \cite{zhang_robust_2013}, or model predictive control \cite{santillan-lemus_optimal_2019} to deal with the high topological complexity and intermittency in this context.
These approaches have in common that they require precise model knowledge and are specifically designed for a certain level of complexity. Adding components with complex behaviours might make such analytical approaches infeasible and model inaccuracies risk impairing control performance in practice. 

Reinforcement learning (RL) controllers overcome these limitations as they do not require explicit model knowledge and can approximate arbitrarily complex control laws without specific adjustments. They have recently become a subject of increased attention in the context of economic dispatch \cite{chen_reinforcement_2022}. 
For example, Yu et al. \cite{yu_deep_2019} investigate deep reinforcement learning for smart home energy management while the study in \cite{wei_wind_2022} focuses on an optimal wind farm bidding model, which makes use of maximum entropy based reinforcement learning for increased robustness. The works in \cite{liu_distributed_2018} and \cite{hao_distributed_2021} propose a control scheme for micro grids based on cooperative reinforcement learning to avoid a central planning entity. Further works improve constraint compliance by utilising constrained reinforcement learning \cite{li_online_2021}, \cite{li_learning_2022}. This provides a probabilistic notion of safety, however, supply reliability is paramount for power systems and the system safety should be certifiable. 

As RL controllers cannot themselves guarantee safety, we safeguard the RL controller's outputs by making use of formal methods. Such methods allow verifying whether given safety specifications are fulfilled. An important line of research in this field uses temporal logic to specify the safety constraints which is, however, only suitable for discrete state and action spaces \cite{alshiekh2018safe}, \cite{hasanbeig2019reinforcement}. Safe sets can furthermore be defined explicitely using control barrier functions \cite{wang_ensuring_2022}, \cite{cheng_end--end_2019}, but finding these functions is difficult for complex systems. Therefore, we follow a different line of research which projects the action of the RL controller into a previously defined safe action space. We use an optimal control formulation \cite{gros_safe_2020}, \cite{wabersich_predictive_2021} to ensure a minimally invasive action projection and constrained zonotopes to define the admissible state and action sets. %The safe action space can for example be defined using control barrier functions \cite{wang_ensuring_2022}, \cite{cheng_end--end_2019}, while we use an optimal control formulation \cite{gros_safe_2020}, \cite{wabersich_predictive_2021} to This minimally invasive action projection minimises the impact of the safety layer on optimality. It can be achieved, for example, via control barrier functions \cite{wang_ensuring_2022}, \cite{cheng_end--end_2019}, while we use an optimal control formulation \cite{gros_safe_2020}, \cite{wabersich_predictive_2021}.

The formal safeguarding requires model knowledge, which seems to relativise the prime advantage of model-free reinforcement learning. However, the models can be significantly simpler, as long as they are an abstraction of reality\footnote{Such a model produces more behaviors than the real system by injecting non-determinism.}, while the capabilities of machine learning can be leveraged for the cost optimisation. \\

This paper develops a provably safe RL control scheme for economic dispatch in which actions of the RL agent are safeguarded by a minimally invasive safety layer. We include a time-dependent constraint, which encodes the islanding contingency \cite{lopes2006defining}. This contingency represents the possibility that all connections to external grids are interrupted for a certain amount of time and is underrepresented in the literature.

Our main contributions are:
\begin{itemize}
    \item Integration of a safety layer guaranteeing the safety of RL controllers in the context of economic dispatch.
    \item Formulation of a minimally invasive action projection strategy, leveraging constrained zonotope set representations for computational efficiency.
    \item Computation of a time-dependent islanding constraint utilising set-based backwards reachability analysis.
\end{itemize}

In order to demonstrate that our approach is fully applicable in future zero-carbon power systems without conventional power plants, our use case only contains renewable sources. After formalising the investigated scenario in Section \ref{section:problem statement}, we describe the safety verification in Section \ref{section:safety verification}. First, we detail the computation of the islanding constraint and proceed with the development of the action projection algorithm. We then test our approach on a residential use case in Section \ref{section: case study} and finally draw our conclusions in Section \ref{section: conclusion}.

\section{PROBLEM STATEMENT} \label{section:problem statement}

\subsection{Setting}

Our micro grid has the following components: A set ${\mathcal{G}}$ of renewable energy generators, a set ${\mathcal{B}}$ of energy storage systems, a set ${\mathcal{D}}$ of residential or industrial consumers, i.e. loads, and a set ${\mathcal{M}}$ of connections to external grids which we call markets as it is assumed that one can buy and sell power at these nodes.

We regard the power network as an undirected graph, in which components are connected through power lines. As we want to focus on high-level planning and safeguard contingency constraints, we only consider active power flows and ignore line losses. Accordingly, every component $i$ can inject active power, i.e. $p_i > 0$, or withdraw active power, i.e. $p_i < 0$. We consider a discrete-time system with sample time interval $\tau$ and control horizon $T$.

Based on the intuition that one would always want to maximise their yield, we regard renewable energy generators as non-dispatchable. Accordingly, the central task is to optimally control the behaviour of the energy storage systems in tune with imports/exports of external energy.

\subsection{Economic dispatch}

Before formalising the economic dispatch problem, we introduce the cost function of energy storage systems as \cite[Sec. II]{malysz_optimal_2014}
\begin{equation}
    C^{\mathcal{B}}_{i}(p_{t, i}) = \tau \sigma_i \ |p_{t, i}|,   
\end{equation}
where $\sigma_i$ is a coefficient based on the equipment cost and expected degradation. The cost function for market nodes is given by \cite[Sec. 3]{santillan-lemus_optimal_2019}
\begin{equation}
    C^{\mathcal{M}}_{t, i}(p_{t, i}) =
    \begin{cases}
      \tau \phi_{t, i}^B p_{t, i}, & \text{if}\ p_{t, i} \geq 0 \\
      \tau (-\phi_{t, i}^S p_{t, i}), & \text{if}\ p_{t, i} < 0,
    \end{cases}
\end{equation}
where $\phi_{t, i}^B$ is the momentary price-per-unit for importing power from the external grid and $\phi_{t, i}^S$ is the price-per-unit for exporting power. We assume our network to be a pure price taker, i.e. to not influence the market prices.

The economic dispatch problem can then be posed as the minimisation of the total system cost over the control horizon
\begin{equation}
     \label{economic_dispatch}
    \min \sum_{t=0}^T \bigg(\sum_{i \in {\mathcal{B}}} C^{\mathcal{B}}_{i}(p_{t, i}) + \sum_{i \in {\mathcal{M}}} C^{\mathcal{M}}_{t, i}(p_{t, i}) \bigg).
\end{equation}

In order to apply reinforcement learning, we model the economic dispatch problem as a \textit{Markov Decision Process (MDP)}, which is described by the tuple $({\mathcal{S}}, \mathcal{A}, T, r, \gamma)$. Here, $\mathcal{S}$ and $\mathcal{A}$ are the MDP state set and action set, respectively, $T(s, a)$ is the transition probability distribution over $\mathcal{S}$ after applying action $a$ in state $s$, and $r(s, a)$ is the reward function returning a scalar reward after applying action $a$ in state $s$. The discount factor $0 \leq \gamma \leq 1$ weights future rewards. Subsequently, we discuss the state and action sets as well as the reward function in detail.

The state of the MDP (also known as observation) consists of the state of charge $e_t$, the power demand $p_t^{\mathcal{D}}$, the power generation $p_t^{\mathcal{G}}$, the market buying price $\phi_t^B$, the market selling price $\phi_t^S$, and forecasts $\hat{p}_t^{\mathcal{D}}, \hat{p}_t^{\mathcal{G}}, \hat{\phi}_t^B, \hat{\phi}_t^S$  for the last four mentioned quantities. We write $s_t~=~[e_t, p_t^{\mathcal{D}}, p_t^{\mathcal{G}}, \phi_t^B, \phi_t^S, \hat{p}_t^{\mathcal{D}}, \hat{p}_t^{\mathcal{G}}, \hat{\phi}_t^B, \hat{\phi}_t^S]^T$, where each forecast can consist of several predictions for different time horizons. Please note that unlike for states of deterministic dynamical systems, forecasts can be integrated beneficially in MDPs.

We expect predictions for a quantity $z$ for some time $t$ in the future as $\hat{z}_{t|t_0} = z_{t} + \xi_{t|t_0}$. Here, $z_{t}$ is the true value, $\hat{z}_{t|t_0}$ is the prediction based on the knowledge at time $t_0$, and $\xi_{t|t_0} \in \Xi_{t|t_0}$ is noise sampled from a bounded disturbance set, the size of which is linearly increasing with the prediction horizon \cite{pinson_on-line_2004}.

Assuming $n$ energy storage systems and $m$ market nodes, the action $a_t = \begin{bmatrix}p_{t, 1}^{\mathcal{B}} &  ... &  p_{t, n}^{\mathcal{B}} &  p_{t, 1}^{\mathcal{M}} &  ... &  p_{t, m}^{\mathcal{M}}\end{bmatrix}^T$ consists of the power set points of all energy storage systems and market nodes.

While the central task of the RL agent is to minimise the operational cost, a notion of safety is helpful to reduce interventions of the safety layer as much as possible. Therefore, we pose the reward function as
\begin{equation}
    \label{reward_fcn}
    r(s_t, a_t) = - \alpha C(s_t, a_t) - \beta r^S_t,
\end{equation}
where $C(s_t, a_t)$ is the system cost from \eqref{economic_dispatch}, $\alpha, \beta \in [0,1]$ are weighting coefficients, and $r_t^S = \|a_t - a_t^S\|$ is a safety correction penalty depending on the safe action $a_t^S$ returned by the safety layer \cite{cheng_end--end_2019}.

\subsection{Dynamic system and constraints}
While the RL agent is model-free, the safety layer models the micro grid as a constrained dynamic system.
The state $x_t \in \mathbb{R}^n$ of the dynamic system is the vector of the charge states
$
   x_t = e_t,
$
while the input $u_t \in \mathbb{R}^{n+m}$ is given by
$
    u_t = \begin{bmatrix}
                    p_t^{\mathcal{B}} &
                    p_t^{\mathcal{M}}
    \end{bmatrix}^T.
$
Let us also introduce $\mathbf{0}$ as an all-zeros matrix and $\mathbf{1}$ as an all-ones matrix with appropriate dimensions, respectively. 

\subsubsection{Dynamics}
The amount of stored energy of any energy storage system $i$ follows the discrete-time dynamic equation (adapted from \cite[Eq. 19]{macedo_optimal_2015})
\begin{equation}
\label{dynamics_ESS}
e_{t+1, i} = e_{t,i} - \tau \ \eta_{t, i} p_{t, i}^{\mathcal{B}}  - \tau \ \mu_i e_{t,i},
\end{equation}
with
\begin{equation*}
    \eta_{t,i} =
    \begin{cases}
      \frac{1}{\eta_i^D}, & \text{if}\ p_{t,i} \geq 0 \\
      \eta_i^C, & \text{if}\ p_{t,i} < 0,
    \end{cases}
\end{equation*}
where $\eta_i^D \in [0, 1]$ is the discharging efficiency, $\eta_i^C \in [0, 1]$ is the charging efficiency, and $\mu_i \in [0, 1]$ is the self-discharge coefficient.
Using the notation introduced in the beginning of this section, we formulate \eqref{dynamics_ESS} as a linear time-variant system of the form
\begin{equation}
\label{linear_system}
    x_{t+1} = A x_t + B_t u_t, 
\end{equation}
where $A \in \mathbb{R}^{n \times n}$ is a diagonal matrix with each diagonal entry $A_{ii} = (1-\mu_i \tau)$ and $B_t = \begin{bmatrix} B_t^{\mathcal{B}} & \mathbf{0}^{n \times m} \end{bmatrix}^T$, where $B_t^{\mathcal{B}} \in \mathbb{R}^{n \times n}$ is a diagonal matrix with each diagonal entry $B_{t, ii}^{\mathcal{B}} = - \eta_{t, i} \tau$ encoding the charge/discharge efficiencies. 

\subsubsection{Power balance constraint}
The amount of power injected must equal the amount of power withdrawn from the network at any time:
\begin{equation}
\label{bus_balance}
    0 = \underbrace{\sum_{i \in {\mathcal{B}}} p_{t, i} + \sum_{i \in {\mathcal{M}}} p_{t, i}}_{h u_t} + \underbrace{\sum_{i \in {\mathcal{D}}} p_{t, i} + \sum_{i \in {\mathcal{G}}} p_{t, i}}_{d_t},
\end{equation}
where $h = \mathbf{1}^{1 \times (n+m)}$ during normal operation and $h = \begin{bmatrix} \mathbf{1}^{1 \times n} & \mathbf{0}^{1 \times m} \end{bmatrix}$ during islanding mode.

\subsubsection{Rate constraint}
For all energy storage systems and markets, the rate of injected/withdrawn power is constrained by
\begin{equation}
    \label{constraints_rate}
    u_t \in \mathcal{U}^{LIM} = [\underline{u}, \overline{u}],
\end{equation}
where $\underline{u}_i~\leq~0$ is the maximum rate of charge, and $\overline{u}_i~\geq~0$ is the maximum rate of discharge in case of an energy storage system. For an external grid connector $\underline{u}_i \leq 0$ is the maximum export rate, and $\overline{u}_i \geq 0$ is the maximum import rate.
We introduce the set of inputs fulfilling \eqref{bus_balance} and \eqref{constraints_rate} as admissible input set 
\begin{equation} \label{admissible input set}
    \mathcal{U}_t^a = \{ u_t \in [\underline{u}, \overline{u}] \mid h u_t + d_t = 0 \}.
\end{equation}

\subsubsection{Storage constraint}
The amount of stored energy in the energy storage systems is constrained by
\begin{equation}
    \label{constraints_storage}
    x_t \in \mathcal{X}^{LIM} = [\underline{x}, \overline{x}],
\end{equation}
where we call $\mathcal{X}^{LIM}$ the admissible state set. 

\subsubsection{Islanding constraint}
The islanding constraint specifies that the system state has to stay in the admissible set for the number of time steps $H$ in islanding mode, i.e. while $p_{t}^{\mathcal{M}} = \mathbf{0}$. We formulate this as a constraint on the current system state
\begin{equation}
    \label{constraints_islanding}
    x_{t} \in \mathcal{X}_{t}^{S} \subseteq \mathcal{X}^{LIM},
\end{equation}
where we call $\mathcal{X}_{t}^{S}$ the safe state set. Please note that we assume our problem to be well-posed, i.e. the dimensioning of the components to be sufficient for the selected islanding horizon $H$. The computation of $\mathcal{X}_{t}^{S}$ is fairly involved and is described in the next section which introduces the formal safeguarding mechanisms in detail.

\section{SAFETY VERIFICATION} \label{section:safety verification}

\subsection{Computation of the safe state set} \label{subsection:scs compuation}

The reachable set $\mathcal{R}_t$ of our system during islanding mode comprises all states that can be reached on the time interval $[t_0, t_0 + t]$, from any state within the initial state set $\mathcal{X}_{t_0}$, for any sequence of admissible inputs, and taking into account load and generation forecasts. Although \eqref{linear_system} is fully decoupled, a component-wise reachability analysis is not possible because the inputs are coupled by the power balance equation.

We call $\mathcal{X}_{t_0}$ safe if it guarantees a non-empty intersection of the reachable set with the admissible state set in each time step during islanding mode: 
$
\mathcal{R}_t \cap \mathcal{X}^{LIM} \neq \emptyset \quad \forall t \in [t_0, t_0+H]
$.
To compute $\mathcal{X}_{t_0}^S$, we first introduce the one-step backwards reachability of the system as (adapted from \cite[Eq. 13]{schurmann_formal_2021})
\begin{equation}
    \mathcal{X}_{t-1} \gets A^{-1} \big( \mathcal{X}_t \oplus (-1) B_{t-1} \mathcal{U}_{t-1}^{aI} \big),
\end{equation}
where $\oplus$ signifies the Minkowski sum\footnote{$\mathcal{A} \oplus \mathcal{B} = \{a + b \ | \ a\in \mathcal{A}, b \in \mathcal{B}\}$.} and $\mathcal{U}^{aI}$ is the admissible input set during islanding mode. As $A$ is a diagonal matrix, the existence of its inverse is guaranteed.

\begin{figure}[t] 
\centering
\includegraphics[width=0.43\textwidth]{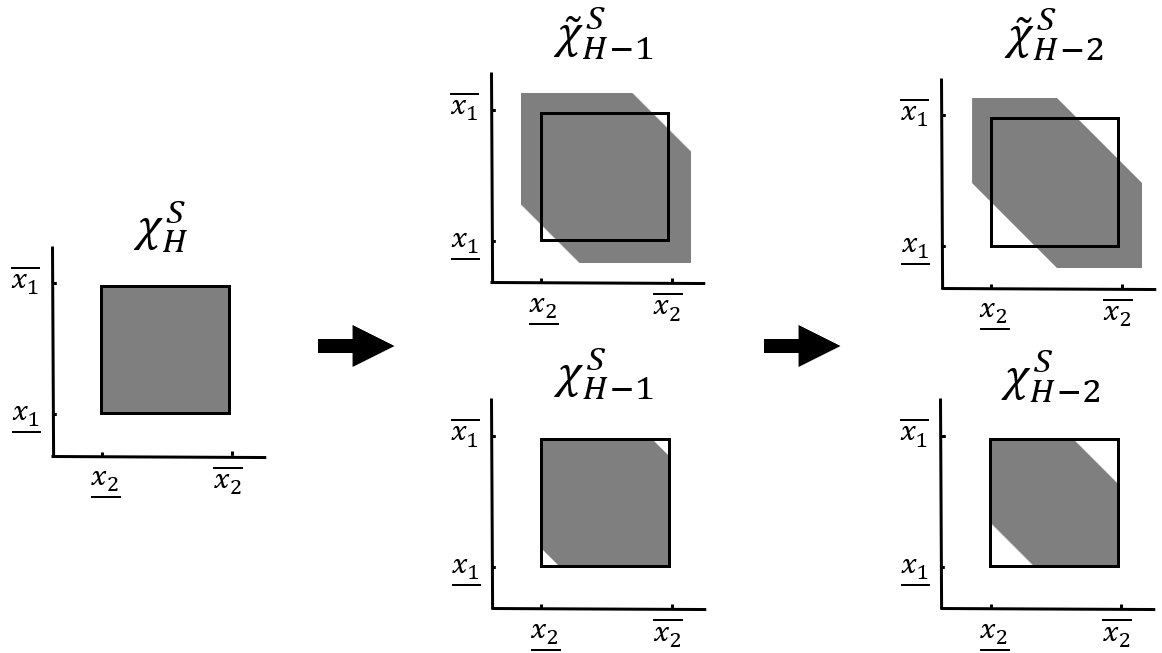}
\caption{Schematic representation of the first two steps of Algorithm \ref{alg:backwards_reachability} for a system with two energy storage systems (computed sets in grey).}
\label{fig:scs_calc}
\end{figure}

We obtain $\mathcal{X}_{t_0}^S$ by evaluating $H$ consecutive one-step backwards reachable sets which are in each step intersected with the admissible state set $\mathcal{X}^{LIM}$ to ensure the storage constraint. The process is formally described in Algorithm \ref{alg:backwards_reachability} and schematically visualised in Fig. \ref{fig:scs_calc}.

\begin{algorithm}
\caption{Obtain safe state set $\mathcal{X}_{t_0}^S$}\label{alg:backwards_reachability}
\begin{algorithmic}
\State $t \gets H$ \;
\State $\mathcal{X}_t^S \gets \mathcal{X}^{LIM}$ \;
\While{$t > t_0$}
    \State $\Tilde{\mathcal{X}}_{t-1}^S \gets A^{-1} \big( \mathcal{X}_t^S \oplus (-1) B_{t-1} \mathcal{U}_{t-1}^{aI} \big)$ \;
    \State $\mathcal{X}_{t-1}^S \gets \Tilde{\mathcal{X}}_{t-1}^S \cap \mathcal{X}^{LIM}$ \;
    \State $t \gets t -1$
\EndWhile
\end{algorithmic}
\end{algorithm}

According to Algorithm \ref{alg:backwards_reachability}, we require a set representation that is closed under linear map, Minkowski sum, and intersection. While \textit{polytopes} are closed under all of the above operations, computing their Minkowski sum has an exponential complexity with respect to the system dimension \cite[Tab. 1]{althoff_set_2021}. 
For this reason, we use \textit{constrained zonotopes}, which have a polynomial complexity with respect to the system dimension for all required operations \cite[Tab. 1]{althoff_set_2021}.
A constrained zonotope is generally defined as \cite[Eq. 9]{SCOTT2016126}
\begin{equation}
    \label{cZ_general}
    \mathcal{Z}_c := \Big\{ c + G \beta \ \Big| \ \|\beta\|_{\infty} \leq 1, F \beta = b \Big\},
\end{equation}
where $c \in \mathbb{R}^k$ is the zonotope center, $G \in \mathbb{R}^{k  \times g}$ is the zonotope generator matrix, $\beta \in \mathbb{R}^g$ is the vector of zonotope factors, and $F \in \mathbb{R}^{q \times g}, \ b \in \mathbb{R}^q$ encode equality constraints. In the following, we will show how to define both $\mathcal{X}^{LIM}$ and $\mathcal{U}_{t}^{aI}$ as constrained zonotopes, starting with the simpler case of $\mathcal{X}^{LIM}$.

For the following paragraphs, let us introduce $\mathrm{diag}(v)$ as a diagonal matrix with $v$ as diagonal entries and $\circ$ as symbol for the Hadamard product. Furthermore, we establish that we can generally rescale a scalar value $y \in [-1, 1]$ to $z \in [\underline{z}, \overline{z}]$ by computing
\begin{equation} \label{rescaling}
    z = 0.5(\overline{z}-\underline{z})y + \underline{z} + 0.5(\overline{z}-\underline{z}).
\end{equation}

The formulation $\mathcal{X}^{LIM} = \mathcal{Z}_c(c, G, \mathbf{0}, 0)$ follows directly from \eqref{rescaling}, yielding
\begin{equation}
\begin{split}\label{cZ_Xlim}
    c &= \underline{x} + 0.5(\overline{x}-\underline{x})\\
    G &= 0.5 \mathrm{diag}(\overline{x}-\underline{x}).
\end{split}
\end{equation}
This structure is equivalent for arbitrary intervals, which we will also leverage for the second definition. 
Before formulating $\mathcal{U}_{t}^{aI}$ as a constrained zonotope, we make two assumptions. To use the available charge as efficiently as possible in islanding mode, we impose the condition that all energy storage system must behave consistently, i.e. they either all charge, $p_t^{\mathcal{B}} \leq 0$, or they all discharge, $p_t^{\mathcal{B}}> 0$, at any point in time. This avoids losses due to unnecessary charging/discharging. 
Furthermore, to plan as conservatively as possible, we consider the lower bound of load-generation predictions $\underline{\hat{d}}_t$ during islanding mode.\footnote{This holds because our system is monotone. Following the argument in \cite[Sec. 8]{angeli_monotone_2003}, a sufficient condition for this is that our system is a Metzler system with $A_{ij} \geq 0 \text{ for all } i \neq j$ and $B_{ij} \geq 0 \text{ for all } i,j$. We can show this by substituting $B_t' = -B_t$ and $u_t'=-u_t$.}

With these assumptions, we obtain $\mathcal{U}_{t}^{aI} = \mathcal{Z}_c(c, G, F, b)$ by selecting $c,G$ to encode the rate limit intervals $u_t \in [0, \overline{u}]$ for the discharging case and $u_t \in [\underline{u}, 0]$ for the charging case, respectively, equivalently to \eqref{cZ_Xlim}. 
We deduce the selection of $F, b$ exemplarily for the discharging case. Expressing the power balance subject to the rescaling described in \eqref{rescaling} results in
\begin{equation}
\begin{split}
    h u_t & = - \hat{\underline{d}_{t}} \\
    h \big(0.5(\overline{u} \circ \beta) + 0.5\overline{u}\big) & = - \hat{\underline{d}_{t}} \\
    0.5(h \circ \overline{u}^T) \beta & = - \hat{\underline{d}_{t}} - 0.5 (h \overline{u}).
\end{split}
\end{equation}
This finally leads us to define for $\mathcal{U}_{t}^{aI}$
\begin{equation} \label{cZ_charging}
\begin{split}
    c & = 0.5 (h^T \circ \underline{\overline{u}}) \\
    G & = 0.5 \mathrm{diag}(h^T \circ \underline{\overline{u}}) \\
    F & =  0.5 (h \circ \underline{\overline{u}}^T) \\
    b & = - \hat{\underline{d}_{t}} - 0.5 h \underline{\overline{u}},
\end{split}
\end{equation}
where 
\begin{equation*}
    \underline{\overline{u}} = 
    \begin{cases}
      \underline{u}, & \text{if}\ \hat{\underline{d}_{t}} \geq 0 \qquad \text{(charging)} \\
      \overline{u}, & \text{if}\ \hat{\underline{d}_{t}} < 0 \qquad \text{(discharging)}.
    \end{cases}
\end{equation*} 

With these definitions of $\mathcal{X}^{LIM}$ and $\mathcal{U}_{t}^{aI}$ we obtain $\mathcal{X}^S_{t_0}$ as a constrained zonotope following Algorithm \ref{alg:backwards_reachability}. We will use this result in the next section in which we develop the action projection mechanism, which ensures constraint compliance through the safety layer.

\subsection{Action projection}

We can find the safe input $u_t$ closest to the proposed action of the agent by solving a model predictive control problem \cite[Sec. 4]{wabersich_predictive_2021}. While all subsequent arguments would apply analogously to a multi-step mechanism, we increase notational clarity by formalizing only the single-step problem
\begin{subequations}
    \label{mpc}
    \begin{align}
        \min_{u_t} \|a_{t} & - u_t\|^2 \\
        \text{subject to } & \notag \\
        \label{mpc:sys}
        x_{t+1} & = A x_{t} + B_t u_t \\
        \label{mpc:ua}
        u_t & \in \mathcal{U}_t^a \\
        \label{mpc:xs}
        x_{t+1} & \in \mathcal{X}_{t+1}^S.
    \end{align}
\end{subequations}
The set containment conditions \eqref{mpc:ua} and \eqref{mpc:xs} generally render this problem non-linear. Therefore, 
to improve computational efficiency, we will recast this optimisation problem as a quadratic program by leveraging the constrained zonotope formulation of $\mathcal{X}_{t_0}^S$.
This requires some adjustments. Let us recall from \eqref{linear_system} that $B_t$ is time-variant and depends on $u_t$. Therefore, considering the entire admissible input set \eqref{mpc:ua} means that we cannot uniquely determine $B_t$.

To evade this issue, we split the control input for energy storage systems into charge and discharge power and obtain $\tilde{u}_t = \begin{bmatrix} p_{t}^{\mathcal{B}D} & p_{t}^{\mathcal{B}C} & p_{t}^{\mathcal{M}} \end{bmatrix}^T$, where $p_{t, i}^{\mathcal{B}D} \geq 0$ is the discharge power, $p_{t, i}^{\mathcal{B}C} \leq 0$ is the charge power, and $p_{t}^{\mathcal{M}}$ is the power of market nodes. 
This lets us reformulate the system dynamics as a linear time-invariant system
\begin{equation}
\label{linear_system_split}
    x_{t+1} = A x_t + \tilde{B} \tilde{u}_t. 
\end{equation}
Here, $\tilde{B} = \begin{bmatrix} \tilde{B}^{\mathcal{B}} & \mathbf{0}^{2n \times m} \end{bmatrix}^T$ with $\tilde{B}^{\mathcal{B}} =  \begin{bmatrix} B^{\mathcal{B}D} & B^{\mathcal{B}C} \end{bmatrix} \in \mathbb{R}^{2n \times 2n}$, where $B^{\mathcal{B}D}$, $B^{\mathcal{B}C}$ are diagonal matrices with diagonal entries $B_{ii}^{\mathcal{B}D} = - \tau \frac{1}{\eta_{i}^D}$ and $B_{ii}^{\mathcal{B}C} = - \tau \eta_{i}^C$, respectively. 

We define the $n \times n$ square identity matrix $\mathbf{I}^n$ and
recast the admissible input set definition \eqref{admissible input set} to correspond to the split input formulation in \eqref{linear_system_split} as 
\begin{equation} \label{admissible input set split}
    \tilde{\mathcal{U}}_t^a = \{ \tilde{u}_t \in \mathbb{R}^y \mid W \tilde{u}_t \leq w, h \tilde{u}_t = -d_t \},
\end{equation}
where $y = 2n + m$. The inequalities $W \tilde{u}_t \leq w$ encode the rate constraints (8) considering the differentiation between charge and discharge and are given by
\begin{equation*}
    W =  \begin{bmatrix}
            \mathbf{I}^n & \mathbf{0}^{n \times n} & \mathbf{0}^{n \times m} \\
             \mathbf{0}^{n \times n} & -\mathbf{I}^n & \mathbf{0}^{n \times m} \\
             -\mathbf{I}^n & \mathbf{0}^{n \times n} & \mathbf{0}^{n \times m} \\
             \mathbf{0}^{n \times n} & \mathbf{I}^n & \mathbf{0}^{n \times m} \\
            \mathbf{0}^{m \times n} & \mathbf{0}^{m \times n} & \mathbf{I}^m  \\
            \mathbf{0}^{m \times n} & \mathbf{0}^{m \times n} & -\mathbf{I}^m
        \end{bmatrix} \text{, }
    w = 
         \begin{bmatrix}
            \overline{p}^\mathcal{B} \\
            - \underline{p}^\mathcal{B} \\
            \mathbf{0}^{n \times 1} \\
            \mathbf{0}^{n \times 1} \\
            \overline{p}^{\mathcal{M}} \\
            - \underline{p}^{\mathcal{M}}
        \end{bmatrix}.
\end{equation*}

We compute $\mathcal{X}_{t+1}^S$ as per Algorithm \ref{alg:backwards_reachability}. The formulation as constrained zonotope allows us to rewrite the safe state set containment condition from \eqref{mpc} as 
\begin{equation} \label{cZ containment reformulation}
\begin{split}
    x_{t+1} \in \mathcal{X}_{t+1}^S \overset{\eqref{cZ_general}, \eqref{linear_system_split}}{\Leftrightarrow} & \, Ax_t + \tilde{B} \tilde{u}_t = c+ G \beta \\
    & \text{ s.t. } F \beta  = b, \, \|\beta\|_{\infty} \leq 1.
\end{split}
\end{equation}

To obtain an optimisation problem with purely linear constraints, we replace \eqref{mpc:sys}, \eqref{mpc:xs} with the formulation in \eqref{cZ containment reformulation} and extract the set properties from \eqref{admissible input set split} to substitute \eqref{mpc:ua}. 
We then concurrently solve for $\tilde{u}_t$ and $\beta$ by defining $\tilde{u}_t' = \begin{bmatrix}\beta & \tilde{u}_t\end{bmatrix}^T$ and compute the safe action $a^S_t = Z \tilde{u}_t'$ by solving 
\begin{equation}
    \label{safety_validation_cZ}
    \begin{split}
        \min_{\tilde{u}_t'} \|a_{t} & - Z \tilde{u}_t'\|^2 \\
        \text{subject to } &\\
        \begin{bmatrix}
                   G & -\Tilde{B} \\
                   F & \mathbf{0}^{q \times y} \\
                   \mathbf{0}^{1 \times g} & \mathbf{1}^{1 \times y}
            \end{bmatrix}
        \tilde{u}_t' & = 
            \begin{bmatrix}
               A x_t - c  \\
               b \\
               - d_t
        \end{bmatrix} \\
        \begin{bmatrix}
               \mathbf{I}^{g} & \mathbf{0}^{g \times y} \\
               - \mathbf{I}^{g} & \mathbf{0}^{g \times y}\\
               \mathbf{0}^{y+m \times g} & W \\
            \end{bmatrix}
        \tilde{u}_t' & \leq 
            \begin{bmatrix}
               \mathbf{1}^{g \times 1} \\
               \mathbf{1}^{g \times 1} \\
               w
        \end{bmatrix},
    \end{split}
\end{equation}
where $Z = 
    \begin{bmatrix}
           \mathbf{0}^{n \times g} & \mathbf{I}^n & \mathbf{I}^n & \mathbf{0}^{n \times m} \\
           \mathbf{0}^{m \times g} & \mathbf{0}^{m \times n} & \mathbf{0}^{m \times n} & \mathbf{I}^m
    \end{bmatrix}$. \\

\section{Case Study} \label{section: case study}
\subsection{Scenario}
Our case study is a household with one photo-voltaic panel, one external grid connection, and two identical batteries. The system is based on the setup in \cite{ioli_smart_2017} including the load and solar panel power generation data, which are based on real-world measurements of 220 days. The system parameters are listed in Table \ref{table:component_parameters}, where power and charge values are given in base units kW and kWh, respectively. The electricity prices are assumed to be constant and represent household buying\footnote{https://strom-report.de/strompreise/strompreisentwicklung/} and selling\footnote{https://revotec-energy.de/eeg-foerderung-photovoltaik/} prices in Germany for 2022. The cost coefficient for the batteries is based on a typical home battery system assuming a life span of 20 years.\footnote{https://www.cleanenergyreviews.info/blog/home-solar-battery-cost-guide}

\begin{table}[t]
\caption{Component parameters}
\label{table:component_parameters}
\begin{center}
\renewcommand{\arraystretch}{1.5}
\begin{tabular}{c c c c c c c c}
\toprule
$\overline{p}^{\mathcal{B}}$ & 3.50 & $\underline{p}^{\mathcal{B}}$ & -3.50 & $\overline{e}^{\mathcal{B}}$ & 6.54 & $\underline{e}^{\mathcal{B}}$ & 0.34 \\
$\eta^D$ & 0.98 & $\eta^C$ & 0.98 & $\mu$ & 0.012 & $\sigma$ & 0.15 \\
\midrule
$\overline{p}^{\mathcal{M}}$ & 5 & $\underline{p}^{\mathcal{M}}$ & -5 & $\phi^S$ & 0.06 & $\phi^B$ & 0.30 \\
\bottomrule
\end{tabular}
\end{center}
\end{table}

%\begin{table}[t]
%    \centering
%    \caption{Evaluation Results (50 Days)}
%    \label{table:evaluation_results}
%    \renewcommand{\arraystretch}{1.5}
%    \begin{tabular}{|c|c|c|c|}
%    \hline
%         & safe agent & safe agent WIS & baseline \\
%         \hline
%         max exec time {[s]} & 0.21 & 0.20 & 0.20 \\ 
%         \hline
%         mean exec time {[s]} & 0.01 & 0.01 & 0.02 \\
%         \hline
%         min SoC {[kWh]} & 0.34 & 0.34 & 0.34 \\
%         \hline
%         max SoC {[kWh]} & 6.47 & 6.50 & 6.47 \\
%         \hline 
%         max safety violation & 6.1e-8 & 0.47 & 0.7 \\
%         \hline
%         mean cost & 0.09 & 0.08 & 0.08 \\
%         \hline
%         mean penalty & 0.08 & 0.10 & 0.11 \\
%         \hline
%    \end{tabular}
%\end{table}
\begin{figure}[t]
    \centering
    \includegraphics[scale=1]{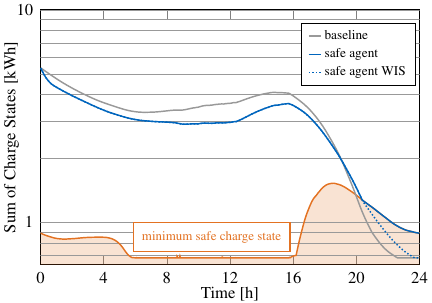}
    \caption{Islanding constraint satisfaction during an exemplary day for baseline agent, safe agent, and safe agent without islanding safeguard.}
    \label{fig:soc_scs}
\end{figure}

The sample time is $\tau = \frac{1}{60}$ (one minute), the control horizon is 24 hours, $T = 1440$, and the islanding horizon is $H=60$ (one hour). For sampling training episodes, 100 full days have been randomly selected and another 20 randomly selected days are used for evaluation. The initial charge states of both batteries are selected randomly from the safe state set at the beginning of each training episode. The agent receives load and demand forecasts $\hat{p}^\mathcal{B}_t, \hat{p}^\mathcal{G}_t$ for $t \in \{t_0+120,t_0+240,t_0+360,t_0+480\}$, respectively. As  \cite{ioli_smart_2017} does not contain forecasts, we generate the predictions by adding smoothed uniform noise to the smoothed real-world measurements. The base amplitude of the uniform noise is $0.05$ for the generation forecasts and $0.01$ for the load forecasts, respectively. In both cases, the amplitude increases linearly with a coefficient of $1.0014$. The smoothing is implemented by two iterations of a moving-average filter with window size 144.

To highlight the effects of the developed safeguarding mechanisms, we compare two agents. The first one which we call \textit{safe agent} is trained with the safety layer as described above. The \textit{baseline agent} on the other hand is trained with a safety layer that only enforces power balance, rate constraints, and keeps the system inside the admissible state set. The safety correction penalty is applied accordingly. For this agent, satisfaction of the islanding constraint is not enforced, however, if the safe state set is violated, an additional penalty is given which is proportional to the violation. The reward weighting coefficients $\alpha, \beta = 0.5$ have been determined empirically and are identical for both agents. For the evaluation we compare the safe and the baseline agent to a third controller. Therein, the safe agent is deployed with the same safety setting as the baseline agent. We call this the safe agent without islanding safeguard (\textit{safe agent WIS}).

\begin{figure*}[t]
\centering
    \begin{tabular}{cc}
        \begin{minipage}{0.4\linewidth}
        \begin{subfigure}{\linewidth}
        \includegraphics[scale=0.8]{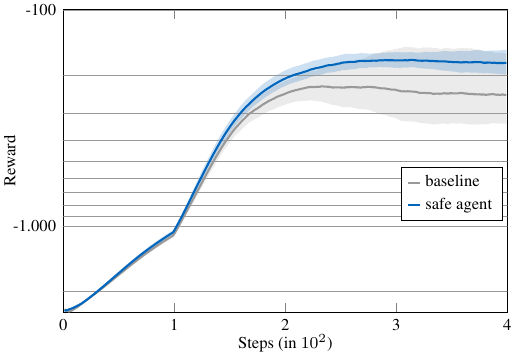}
        \caption{Total reward}
        \label{fig:reward_training_baseline_vs_safe}
        \end{subfigure}
        \end{minipage} &
        \begin{minipage}{0.4\linewidth}
        \begin{subfigure}{\linewidth}
        \includegraphics[scale=.8]{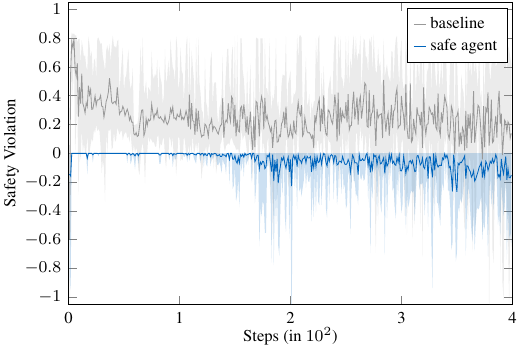}
        \caption{Maximum safety violation}
        \label{fig:scs_training_baseline_vs_safe}
        \end{subfigure}
        \end{minipage}
        \\ 
        \begin{minipage}{0.36\linewidth}
        \begin{subfigure}{\linewidth}
        \vspace{0.3cm}
        \includegraphics[scale=.8]{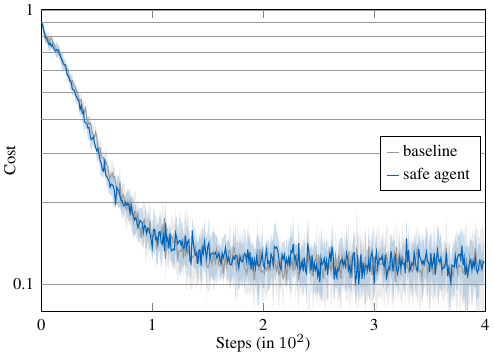}
        \caption{Mean cost}
        \label{fig:cost_training_baseline_vs_safe}
        \end{subfigure}
        \end{minipage} & 
        \begin{minipage}{0.36\linewidth}
        \begin{subfigure}{\linewidth}
        \vspace{0.3cm}
        \includegraphics[scale=.8]{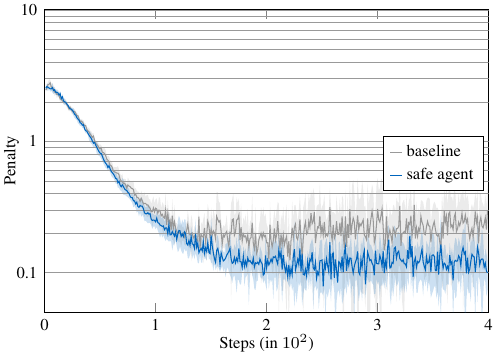}
        \caption{Mean penalty}
        \label{fig:penalty_training_baseline_cs_safe}
        \end{subfigure}
        \end{minipage}
    \end{tabular}
	\caption{Reward, safety violation, cost and penalty curves for training. Logarithmic scaling is used in all plots with horizontal lines.}
	\label{fig:training_baseline_vs_safe}
\end{figure*}

Training is performed in Python 3.8 by utilising \textit{Proximal Policy Optimisation (PPO)} \cite{schulman_proximal_2017} within the framework \textit{Stable-Baselines3} \cite{stable-baselines3}, while the safety layer and power system itself are simulated in Matlab 2021b. The PPO algorithm is run with default hyperparameters except for the batch size, which equals $24$. Both the policy and the value function network have two hidden layers with $32$ neurons each.

\subsection{Results and Discussion}

Fig. \ref{fig:training_baseline_vs_safe} exemplifies the behaviour of the safe and baseline agent during training. We observe that the safe agent converges to a higher mean reward. Fig. \ref{fig:cost_training_baseline_vs_safe} shows that the  evolution of the average cost during training is similar for both agents. The lower performance of the baseline agent can be attributed to the higher average penalty it receives as a result of repeatedly violating the islanding constraint. This is illustrated in Fig. \ref{fig:penalty_training_baseline_cs_safe}. To obtain a scalar indicator for constraint compliance, we use the sum of charge states of the two batteries. We define safety violation as the sum of the minimum charge states contained in the safe state set subtracted by the sum of actual charge states. If the value of the safety violation is positive, satisfaction of the islanding constraint is no longer given. Fig. \ref{fig:scs_training_baseline_vs_safe} displays the maximum safety violation during training for both agents. We observe that the safe agent always satisfies the islanding constraint, while the baseline agent still violates it even after the reward has converged. Furthermore, Fig. \ref{fig:scs_training_baseline_vs_safe} shows that the mean safety violation decreases during training as the agents learn to avoid the respective penalty. In summary, using the full safety layer during training is not only crucial for constraint satisfaction, it is also benefits the performance. 

%\begin{figure*}[t]
%    \subfloat[][Total reward]{
%    \includegraphics[scale=1]{graphics/rewardTraining.pdf}
%    \label{fig:reward_training_baseline_vs_safe}
%    }
%    \subfloat[][Maximum safety violation]{
%    \includegraphics[scale=1]{graphics/scsTraining.pdf}
%    \label{fig:scs_training_baseline_vs_safe}
%    }
%    \subfloat[][Mean cost]{
%    \includegraphics[scale=1]{graphics/costTraining.pdf}
%    \label{fig:cost_training_baseline_vs_safe}
%    }
%	\caption{Reward, safety violation, and cost curves for training.}
%	\label{fig:training_baseline_vs_safe}
%\end{figure*}

For deployment of the trained agents, we additionally investigate the performance of the safe agent without islanding safeguard. Fig. \ref{fig:soc_scs} compares the behaviour of the three controllers on a randomly selected day from the evaluation set. Only the safe agent in combination with the full safety layer fulfills all safety requirements. Without the islanding safeguard, the safe agent violates the safe charge state. However, this violation is smaller compared to the baseline agent. While Fig. \ref{fig:soc_scs} shows one exemplary day, Table \ref{table:evaluation_results} lists the results for $50$ simulated days, where the charge values and execution times are given in kWh and seconds, respectively. 
First, we notice that the minimum and maximum charges of the single batteries (see Table \ref{table:component_parameters}) are never violated. 

Second, we observe that only the the safe agent paired with the safety layer remains within the safe state set over all simulated examples. The maximum safety violation is positive, but in the order of $10^{-8}$. This can be attributed to the constraint tolerance of the solver used for the quadratic program \eqref{safety_validation_cZ}, which is set to $10^{-8}$. 

Third, it becomes clear that the safe agent without the islanding safeguard performs slightly better than the baseline agent with respect to the maximum safety violation and the mean penalty. This implies that training the agent with the full safety layer is beneficial even if the safety layer is not active during deployment. 

A fourth notable result is the maximum execution time of the safety layer. For a sample time of one minute, the maximum compute time over all three combinations is $0.23$ seconds (Intel(R) Core(TM) i7-11800H, 2.3GHz), rendering the safety mechanism real-time capable.

\begin{table}[t]
    \centering
    \caption{Evaluation results (50 Days)}
    \label{table:evaluation_results}
    \renewcommand{\arraystretch}{1.5}
    \begin{tabular}{lccc}
    \toprule
         & safe agent & safe agent WIS & baseline \\
         \midrule
         max exec time & 0.23 & 0.20 & 0.20 \\ 
         mean exec time & 0.02 & 0.01 & 0.02 \\
         min charge state & 0.34 & 0.34 & 0.34 \\
         max charge state & 6.30 & 6.30 & 6.35 \\
         max safety violation & 6.10e-8 & 0.40 & 0.67 \\
         mean cost/day & 122.69 & 121.39 & 114.34 \\
         mean penalty/day & 120.96 & 128.16 & 156.10 \\
    \bottomrule
    \end{tabular}
\end{table}

\section{Conclusion} \label{section: conclusion}

This paper presents a provably safe RL controller for economic dispatch by inclusion of a formal safety layer. Within our framework, constraint satisfaction is certifiable, whereas previous approaches \cite{li_online_2021}, \cite{li_learning_2022} only provide a probabilistic notion of safety. Furthermore, we extend the conventional set of system constraints by a constraint encoding the islanding contingency. We provide an algorithm to calculate this time dependent island constraint utilising set-based backwards reachability analysis. The safe state set is defined as a constrained zonotope, a set representation that has polynomial complexity for all required set operations. The safety layer ensures constraint compliance by projecting the agent's proposed action into the safe action set. To minimise interference, this projection is formulated as a quadratic program. We demonstrate in a real-world-measurement-based residential case study that the controller never violates constraints. The computational efficiency of our approach enables real-time capability of the safety layer. Future studies will generalise the approach by including reactive power, power flow equations, and more complex models for demand and generation forecasts.

%%%%%%%%%%%%%%%%%%%%%%%%%%%%%%%%%%%%%%%%%%%%%%%%%%%%%%%%%%%%%%%%%%%%%%%%%%%%%%%%

\addtolength{\textheight}{-3cm}   % This command serves to balance the column lengths

%%%%%%%%%%%%%%%%%%%%%%%%%%%%%%%%%%%%%%%%%%%%%%%%%%%%%%%%%%%%%%%%%%%%%%%%%%%%%%%%

\bibliography{references}
\bibliographystyle{ieeetr}

\end{document}